\documentclass[10pt,prd,twocolumn,
  nofootinbib,noshowkeys,noshowpacs,
  superscriptaddress,floatfix]{revtex4-2}

\usepackage{amsmath,amsfonts,amsthm,amssymb,tensor,physics,bigints,scalerel,nicefrac}
\usepackage{bbold} 

\usepackage{graphicx} 
\usepackage[usenames,dvipsnames]{color}
\definecolor{darkblue}{RGB}{0,0,196}
\definecolor{darkgreen}{RGB}{0,120,0}

\usepackage{tikz}
\usetikzlibrary{3d,calc,arrows.meta,decorations.markings}
\usepackage{tikz-feynman}
\tikzfeynmanset{compat=1.1.0}

\usepackage{soul}        
\usepackage{marginnote}  
\usepackage[normalem]{ulem} 
\usepackage{cancel}      
\usepackage{comment}     
\usepackage{verbatim}    

\usepackage{caption}
\usepackage{subcaption}

\usepackage{multirow}

\usepackage{stix} 

\usepackage{nameref} 
\usepackage[colorlinks=true,linkcolor=blue,citecolor=blue,urlcolor=blue]{hyperref}

\usepackage{appendix}

\begin{document}
\preprint{}

\title{$B \to D^{(*)} \tau \nu_{\tau}$ decay properties with RIQ model}
\author{Sonali Patnaik}
\email{sonali\_patnaik@niser.ac.in}
\affiliation{National Institute of Science Education and Research, An OCC of Homi Bhabha National Institute, Bhubaneswar, Odisha, India}

\author{Lopamudra Nayak}
\email{lopalmnayak@niser.ac.in}
\affiliation{National Institute of Science Education and Research, An OCC of Homi Bhabha National Institute, Bhubaneswar, Odisha, India}

\author{Sanjay Kumar Swain}
\email{sanjay@niser.ac.in}
\affiliation{National Institute of Science Education and Research, An OCC of Homi Bhabha National Institute, Bhubaneswar, Odisha, India}
\begin{abstract}
In this work we compute the branching fraction of $B \to D^{(*)} \,\tau \, \nu_{\tau}$ and $B_s \to D_s^{(*)}\, \tau\, \nu_{\tau}$ within the \emph{Relativistic Independent Quark Model}, emphasizing the harmonic potential model-dependent analysis of these decay channels in the precision flavor physics era. Considering the experimental observation of longitudinal $\tau$-polarization and fraction of longitudinal polarization at LHCb and Belle, we have also investigated these observables within our model framework which are aligning well with the standard model expectations. We perform a comprehensive analysis of the form factors across the whole accessible kinematic range of $q^2$. Our results are consistent and compatible with other theoretical approaches as well as with the experimental measurements. Furthermore, we evaluated the clean ratios of $B_s$ to $B_0$ in the semimuonic mode that are in accordance with the LHCb measurements, and support the validity of the SU(3) flavor symmetry. Although the concept of new physics is facing a significant challenge at the current TeV scale, semileptonic $B$ decays, nevertheless, always serve as a valuable probe to study the decay dynamics.   
\end{abstract}

\date{\today}




\maketitle
\section{Introduction}
\label{sec:intro}
The weak decays of $B$ and $B_s$, which contain bottom ($b$) quark, and the spectator quarks $q = u, d,s$ are considered promising avenues for uncovering potential signs of physics beyond the Standard Model (BSM). Flavour-changing transitions in $B$-meson decays have attracted significant attention due to several observed discrepancies between experimental data and Standard Model (SM) predictions~\cite{Belle:2009zue,BaBar:2012obs,Belle:2015qfa,LHCb:2017smo,LHCb:2020cyw,CMS:2024syx,LHCb24:2024LHCb}. These anomalies highlight the importance of refining theoretical predictions within the SM framework, particularly through approaches and processes where uncertainties can be tightly controlled. The leading $B$ decays to final meson states with $\tau$ leptons have been observed by BaBar~\cite{BaBar:2012obs,BaBar:2013mob}, Belle~\cite{Belle:2016dyj,Belle:2019oag,Belle-II:2025yjp}, and LHCb~\cite{LHCb:2015gmp,LHCb:2019efc,LHCb:2021lvy,LHCb:2021trn,LHCb:2021xxq,LHCb:2021zwz}, with hints of lepton flavor universality violation (LFUV) - the universality of the electroweak gauge couplings across the three established generations of leptons. Recent measurements in leptonic and semileptonic decays, such as $B \to \tau \nu_{\tau}$ and $B \to D\,(D^*)\, \ell\, \nu_{\ell}$ with $\ell = e, \mu$ or $\tau$ challenge the SM expectation of lepton universality. Confirming LFUV and identifying possible new interactions could offer valuable clues about physics BSM, including dark matter, matter-antimatter asymmetry, and electroweak dynamics. In view of these, the  anomalies in semileptonic $B$ meson decays can be interpreted as potential signs of new physics (NP). On the other hand, it is equally valid to recognize that the idea of NP at the TeV scale is under considerable pressure, given the lack of observed deviations from the SM in high energy experimental searches. While the significance of each deviation on its own is not particularly strong, collectively, they present a compelling and intriguing pattern.

The Belle II Collaboration~\cite{Belle-II:2025yjp} recently reported updated measurements of LFUV observables using a 365 fb$^{-1}$ data set collected at the $\Upsilon(4S)$ resonance. By reconstructing the companion $B$  meson in semileptonic modes and the $\tau$ lepton via its leptonic decays, they obtained the following result: 

\begin{center}
$\mathcal{R}_{D} = 0.418 \pm 0.074 \,(stat.)\, \pm 0.051 \,(syst.)$,
\end{center}

\begin{center}
$\mathcal{R}_{D^*} = 0.306 \pm 0.034\, (stat.) \pm 0.018 \,(syst.)$.
\end{center}
These results are consistent with the world average values and differ from the SM expectations by a combined significance of 1.7 $\sigma$. Meanwhile, LHCb~\cite{LHCb:2024jll} presented the same measurements:

\begin{center}
$\mathcal{R}_{D} = 0.249 \pm 0.043 \pm 0.047$,
\end{center}
\begin{center}
$\mathcal{R}_{D^*} = 0.402 \pm 0.081 \pm 0.085$,
\end{center}

\noindent which are in agreement with previous results and contribute to the global averages~\cite{HFAG:afp}

\begin{center}
$\mathcal{R}_{D} = 0.342 \pm 0.026$,
\end{center}
\begin{center}
$\mathcal{R}_{D^*} = 0.287 \pm 0.012$.
\end{center}

These averages are obtained by considering correlations among systematic uncertainties and using the COCO averaging tool. The updated measurements from Belle II and LHCb continue to provide important tests of lepton flavor universality and help refine the global understanding of semileptonic $B$ meson decays. However, the world average is gradually approaching SM predictions~\cite{Iguro:2020cpg, Bordone:2019vic}, and the significance of the deviation remains above 3$\sigma$, thanks to reduced uncertainties. That said, persistent anomalies in differential branching fractions and angular analyses of the muon modes suggest ongoing discrepancies. The theoretical and experimental measurements will contribute to a more dedicated parameterization of form factors and their accurate determination involving in semileptonic transitions. This provides strong motivation for a timely investigation of semileptonic $B$ channels.

The Belle measurement of longitudinal $\tau$ polarization, $P_{\tau}(D^*)$~\cite{Belle:2016dyj, Belle:2019ewo} and the fraction of $D^*$ longitudinal polarization, $F_L^{D^*}$, for $B \rightarrow D^* \tau \nu_{\tau}$ have attracted significant attention for investigating these observables.
\begin{eqnarray}
    P_{\tau}(D^*) &=& - 0.38 \pm 0.51 \,({\rm stat.})^{+0.21}_{-0.16}\, ({\rm syst.})\,,\nonumber\\
    F_{L}(D^*) &=& 0.60 \pm 0.08 \,({\rm stat.}) \pm (0.04)\,({\rm syst.})\,,
\end{eqnarray}
whereas, the SM predictions of these observables are~\cite{Tanaka:2012nw,Huang:2018nnq,Bhattacharya:2018kig}
\begin{eqnarray}
P_{\tau}(D^*) &=& - 0.497 \pm 0.013\,,\nonumber\\
F_L(D^*) &=& 0.441 \pm 0.006 \quad {\rm or} \quad 0.457 \pm 0.010\,.
\end{eqnarray}
The experimental findings align with the predictions of the SM. Recently, the LHCb~\cite{LHCb:2023ssl} measured the longitudinal polarization fraction of the $D^*$ meson in decays where $B^0 \rightarrow D^{-} \tau^+ \nu_\tau$, with the $\tau$ lepton decaying into three charged pions and a neutrino. This measurement was conducted using proton-proton ($pp$) collision data collected at center-of-mass energies of 7, 8, and 13 TeV, corresponding to an integrated luminosity of 5 fb$^{-1}$. For the lower and higher $q^2$ regions, the measured $D^*$ polarization fractions respectively are
\begin{eqnarray}
F_L^{D^*} = 0.51 \pm 0.07 \pm 0.03  \quad \text{and} \quad
0.35 \pm 0.08 \pm 0.02 \,.
\end{eqnarray}
And the average value over the entire $q^2$ range is
\begin{eqnarray}
F_L^{D^*} = 0.41 \pm 0.06 \pm 0.03\,.
\end{eqnarray}
These outcomes are consistent with both SM observations and findings from the Belle experiment. 

LHCb collaboration~\cite{LHCb:2020hpv} have also lately measured the shape of the differential decay rate for $B^0_s \to D_s^{*} \mu \nu_{\mu}$ as a function of the hadron recoil parameter, corresponding to an integrated luminosity of 1.7 fb$^{-1}$. Measurements of the branching ratio for $B_s \to D_s^{(*)} \mu \nu_{\mu}$ are given to be ($2.31 \pm 0.21$ \%) and ($5.4 \pm 1.1$ \%) respectively~\cite{ParticleDataGroup:2024cfk}. However, the decays involving $\tau$ mode have not been measured yet. It is expected that in near future, more
and more $B_s$ channels will be precisely measured experimentally. 

Theoretically, the decay channels $B_s \to D_s^{(*)} \ell^+ \nu_{\ell^-}$ and $B \to D^{(*)} \ell^+ \nu_{\ell^-}$ are driven by the same quark-level transition $b \to c \ell \nu_{\ell}$, which differs only by the spectator quark, which can be a $u , d$ or $s$ quark. These processes are connected through SU(3) flavor symmetry, under which the semileptonic decay characteristics are expected to exhibit close similarity. If the anomalies currently observed in $B \to D^{(*)} \ell \nu$ decays are caused by contributions from NP, corresponding signatures should also be present in $B_s \to D_s^{(*)} \ell \nu$ decays. Therefore it is both important and compelling to investigate $B_s \to D_s^{(*)} \ell \nu$ channel as well and measure the key observables involving such as $P_\tau(D_s^{(*)})$, and  $F_L(D_s^*)$ to check for analogous deviations. These angular observables serve as essential tools for constraining and identifying the most viable BSM theories. In $b\rightarrow c \tau \nu_{\tau}$ decays, the subsequent  decay of the $\tau$ lepton within the detector permits the extraction of its polarization fraction along a specific polarization axis. This polarization observable being hadronic final state dependent, exhibits pronounced sensitivity to BSM effects, thereby offering a complementary handle alongside branching ratios and kinematic distributions in three-body decays. These considerations motivate us to provide theoretical predictions for such observables in the present analysis.

In the theory side the study of weak decays deals with crucial significance in calculating the transition form factors. To determine the form factors associated with a given process, one must employ non-perturbative methods such as the Bethe–Salpeter (B-S) equation, QCD sum rules (QCDSRs), or Lattice QCD (LQCD). Specifically, for weak transitions, $B_s \to D_s^{(*)} \ell \nu_{\ell}$ various studies have been conducted. Early analyses largely utilized the well-known Bauer–Stech–Wirbel (BSW) model \cite{Kramer:1992xr,Wirbel:1988ft,Bijnens:1992np,Wirbel:1985ji}. Subsequent works, such as~\cite{Blasi:1993fi,Azizi:2008vt}, applied QCDSRs for their computations. In~\cite{Li:2010bb}, the form factors were evaluated using the covariant light-front quark model (CLFQM). Meanwhile, the authors of~\cite{Li:2009wq} employed light-cone sum rules (LCSRs) to study the form factors at large recoil and used heavy quark effective theory (HQET) to describe the small recoil region. Extensive investigations of $B \to D^{(*)} \ell \nu_{\ell}$ transitions have also been studied using the Covariant Quark Model (CQM), a relativistic constituent quark model in which mesons are treated as bound states via an effective Lagrangian~\cite{Ivanov:2015tru,Ivanov:2016qtw,Ivanov:2017mrj}. The model features Lorentz-covariant quark currents with Gaussian vertex functions and computes the full $q^2$ dependence of form factors from one-loop quark diagrams, without relying on heavy-quark expansions. LQCD provides SM form factors results at zero recoil for both $B \to D$ and $B_{s} \to D_{s}$~\cite{MILC:2015uhg,McLean:2019qcx,Harrison:2017fmw}. Results for $B_{s} \to D_{s}^*$ and $B \to D^*$ transitions in the whole $q^2$ range have recently emerged~\cite{Harrison:2023dzh,Penalva:2023snz,Blossier:2021azl,FermilabLattice:2021bxu,Aoki:2023qpa}.   LQCD predictions for differential decay rates are now possible, and by combining HQET and QCD sum rules, estimates for $B \to D \, \tau\, \nu_{\tau}$ can be made, though with less precision compared to data-driven methods~\cite{Bernlochner:2021vlv}. Also various approaches using model independent analysis have also been established to understand these scenarios~\cite{Tanaka:2012nw, Watanabe:2017mip, Bhattacharya:2018kig}.

In parlance, QCD inspired quark models are deemed successful when it can adequately replicate the available experimental data across different hadron sectors. Independent of the specific Lorentz structure of the interaction potential, a phenomenological model gains credibility if it effectively captures the constituent level dynamics within hadrons and can predict a range of hadronic properties including decay behaviors. The ongoing and upcoming experiments in the $B$ sector along with existing theoretical predictions, inspire us to investigate the exclusive semileptonic decays of $b$-flavored mesons using relativistic independent quark (RIQ) model~\footnote{This model has been successfully applied to study static properties of hadrons \cite{Barik:1986mq,Barik:1987zb,Barik:1993aw}, as well as their decay properties, including radiative, weak radiative, rare radiative decays \cite{Barik:1994vd,Priyadarsini:2016tiu}; leptonic, weak leptonic, radiative leptonic decays \cite{Barik:1993yj,Barik:1993aw}; and non-leptonic decays \cite{Barik:2001vp,Kar:2013fna,Nayak:2022qaq}.Moreover, RIQ model has also been widely extended to study semileptonic decays of heavily flavored mesons \cite{Barik:1996xf,Barik:2009zz}.}. To strengthen our model reliability it is essential to expand the applicability of a quark model to investigate the decay channels: $B \to D\,(D^*)\, \tau\, \nu_{\tau}$ and $B_s \to D_s\,(D_s^*) \,\tau \,\nu_{\tau}$, specifically the $\tau$ modes.  These decays are in fact, governed by six independent invariant hadronic form factors $f_+(q^2)$, $f_-(q^2)$ , $V(q^2)$ , $A_0(q^2)$ , $A_+(q^2)$ \& $A_-(q^2)$. In the RIQ model framework, the invariant weak decay form factors formulated as overlap integrals of meson wave functions have been extracted in several decay channels~\cite{Patnaik:2019jho,Nayak:2021djn,Patnaik:2022moy,Nayak:2024esq}, showing good agreement with experimental data and other SM predictions. In this article, we calculate the form factors in the entire kinematic range: $0 \le q^2 \le (M_B - M_{D_{(s)}}^{(*)})^2$, along with it, the branching fraction predictions for the decays involving $\tau$ mode. Similar to the initial measurement of $P_{\tau}^{D^*}$ at Belle \cite{Belle:2016dyj, Belle:2019ewo}, and $F_L(D^*)$ at LHCb~\cite{LHCb:2023ssl}, we also give similar predictions for the observables within the SM framework, using our RIQ model conventions.

The article's structure is as follows: Section~\ref{sec:physical_obs} defines the physical observables used throughout the study. Sections~\ref{sec:RIQ} and~\ref{sec:num_res} describe our theoretical framework for $b \to c$ transitions and present the resulting analyses and findings, respectively. Lastly, Section~\ref{sec:conclusions} summarizes the main results and discussions.
 
\section{Physical Observables}
\label{sec:physical_obs}
Here, we outline the set of observables that will be employed in our subsequent analysis and phenomenological discussion of $b \rightarrow c$ transitions.

\begin{itemize}
    \item \emph{Branching fraction}: The first observable is the total branching fraction which is most commonly considered in experimental searches and is given by
    \begin{equation}
    Br_{\rm tot} = \int_{m_l^2}^{(M-m)^2} \Bigg(\frac{d\mathcal{B}(q^2)}{dq^2}\Bigg)\;dq^2\,,
    \end{equation}
  which can be computed using \eqref{eq:dw} in \ref{sub:helicity}.

\item \emph{Lepton-polarization asymmetry}: A study of the decay to the charged lepton with a specific polarization state allows one to measure the lepton-polarization asymmetry defined as:
    \begin{equation}
P_L = \frac{\frac{d\tilde{\Gamma}_i}{dq^2}^{h=-1} - \frac{d\tilde{\Gamma}_i}{dq^2}^{h=+1}}{\frac{d\tilde{\Gamma}_i}{dq^2}^{h=-1} + \frac{d\tilde{\Gamma}_i}{dq^2}^{h=+1}}\,.
\end{equation}
The differential decay rates ${d \tilde{\Gamma}_i}/{dq^2}$ is assosciated with a flip factor $m_{\tau}^2 / (2q^2)$. Here, $h$ represents helicity and is defined as $h = s \cdot p = \pm 1$, where the sign corresponds to the case when the spin polarization is parallel (+) or anti-parallel (-) to the lepton's ($e, \mu, \tau$) momentum direction.  This formulation is consistent with the conventional treatment of polarization asymmetry, which is expressed through the helicity amplitude couplings as described below.
\begin{equation}
P_L = \frac{\tilde{U} + \tilde{L} + \tilde{S} - U - L}{\tilde{U} + \tilde{L} + \tilde{S} + U + L}\,.
\end{equation}
The usual meanings of individual helicity functions has been detailed in~\ref{sub:helicity}

\item \emph{Fraction of Longitudinal Polarization ($F_L$)}: In semitauonic decays of $B$ and $B_s$ the final state vector ($V$) mesons ($D^*$ and $D_s^*$) can be produced with different polarization states -- longitudinal or transverse. The longitudinal polarization fraction denoted as $F_L$ quantifies the proportion of decays in which the vector meson is longitudinally polarized and is given as
\begin{equation}
    F_L^V = \frac{\Gamma^{\lambda_{V} = 0}}{\Gamma^{\lambda_{V} = 0} + \Gamma^{\lambda_{V} = 1} + \Gamma^{\lambda_{V} = -1}}.
\end{equation}
Here $\Gamma^{\lambda_V = 0}$ is the longitudinal decay rate which corresponds to the contribution from the longitudinal helicity structure function $H_L$. $\Gamma^{\lambda_V = \pm1}$is the transverse decay rate which corresponds to the contributions from the transverse helicity structure functions $H_U$, which have been mentioned in~\ref{sub:helicity}.
\end{itemize}
\section{RIQ model framework}
\label{sec:RIQ}
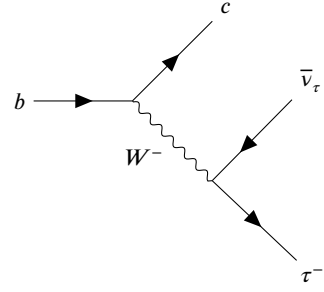
\begin{figure}[t]
\begin{tikzpicture}
  \begin{feynman}
    \vertex (a) {\(b\)};
    \vertex [right=of a] (b);
    \vertex [above right=of b] (f1) {\(c\)};
    \vertex [below right=of b] (c);
    \vertex [above right=of c] (f2) {\(\overline \nu_{\tau}\)};
    \vertex [below right=of c] (f3) {\(\tau^{-}\)};

    \diagram* {
      (a) -- [fermion] (b) -- [fermion] (f1),
      (b) -- [boson, edge label'=\(W^{-}\)] (c),
      (c) -- [anti fermion] (f2),
      (c) -- [fermion] (f3),
    };
  \end{feynman}
\end{tikzpicture}
\caption{quark level transition for $B\, (B_s) \to D\, (D_s^{(*)})\, \tau \,\nu_{\tau}$}
\label{Fig:LOFD}
\end{figure}
 
Studying exclusive semileptonic decays involving non-perturbative hadronic matrix elements can be formulated using a model dependent framework. In this context, we provide a concise theoretical predictions adopting the RIQ model (RIQM) framework. The RIQM is grounded in a confining harmonic potential in the equally mixed scalar-vector form~\cite{Patnaik:2017cbl,Patnaik:2018sym}
\begin{equation}
U(r)=\frac{1}{2}\left(1+\gamma^0\right)\,V(r)\,,
\label{eq:harmonicpotential}
\end{equation}
\begin{center}
    and $V(r)=(ar^2+V_0)$.
\end{center}

Here, $r$ represents a state dependent length parameter, $\gamma^0$ denotes the time-like Hermitian matrix, and the parameters $a$ and $V_0$ are associated with the potential. These parameters were initially established through applications of the model in hadron spectroscopy and in this work it has been  recalibrated to enhance the accuracy of the results. The confining interaction is believed to provide phenomenologically the zeroth order quark dynamics inside the hadron-core through the quark Lagrangian density,
\begin{equation}
{\cal L}^{0}_{q}(x)={\bar \psi}_{q}(x)\;\left[\frac{i}{2}\,\gamma^{\mu}
\partial _\mu - m_q-U(r)\;\right]\;\psi _{q}(x)\,,
\end{equation}
leading to the Dirac equation for individual quark
\begin{equation}
\left[\gamma^0E_q-{\vec \gamma}.{\vec p}-m_q-U(r)\right]\psi_q(\vec r) = 0\,,
\end{equation}
with $\psi_q(\vec r)$ representing the four-component Dirac normalized wave function~\cite{Nayak:2021djn,Patnaik:2023efe}.

The RIQM is a QCD-inspired phenomenological model, akin to other potential models. It characterizes the confinement of constituent quarks within hadrons through an interaction potential with a specified Lorentz structure. This model aims to derive observable properties of composite hadrons from the constituent quark dynamics as direct derivation from first principles of QCD is currently challenging due to inherent complexities. The chosen potential, \eqref{eq:harmonicpotential}, is assumed to depict non-perturbative multi-gluon interactions. Additionally, residual interactions, such as quark-pion coupling from chiral symmetry restoration in the SU(2) sector and one-gluon exchange at short distances, are treated perturbatively in this model to predict the hadron mass spectra~\cite{Barik:1986mq,Barik:1987zb}.The selected confining potential yields a simple and manageable form, facilitating the analysis of various hadronic properties and providing an adequate tree-level description for various decays. To achieve this goal, we expand the RIQM framework's scope to different sectors, demonstrating it as a viable phenomenological scheme for studying various hadronic phenomena. We compare our results with other theoretical approaches and available experimental data. 
\subsection{Theoretical Formalism}
\label{subsec:TF}
The invariant transition matrix element for $B\to D\,(D^*)\, \tau \,\nu_{\tau}$ and  $B_s\to D_s\,(D_s^*) \,\tau \, \nu_{\tau}$ governing by the quark level transition $b \to c \tau \nu_{\tau}$ as depicted in Fig.~\ref{Fig:LOFD} is expressed as:

\begin{equation}
\mathcal{M}(p,k,k_{\tau},k_{\nu_{\tau}})={\frac{\cal G_F}{\sqrt{2}}}V_{cb}\,{\cal H}_\mu(p,k) \,{\cal L}^\mu(k_{\tau},k_{\nu_{\tau}})\,,
\label{eq:TME}
\end{equation}

with ${\cal G}_F$ being the effective Fermi coupling constant, $V_{cb}$ being the  relevant CKM parameter, ${\cal L}^\mu$ and ${\cal H}_\mu$ are leptonic and hadronic current, respectively. Here, $p,k,k_{\tau},k_{\nu_{\tau}}$ denote four momentum of parent ($B$ or $B_s$) and daughter $X$, where $X$ = $D (D^*)$ or $D_s(D_s^*)$ mesons, $\tau$-lepton and neutrino, respectively. 
In the context of decay processes, the physical occurrence relies on the participating mesons being in their momentum eigenstates. Thus, a field-theoretic depiction of any decay process necessitates representing meson bound states through suitable momentum wave packets, reflecting the momentum and spin distribution between the constituent quark and antiquark within the meson core. In the RIQ model approach, the wave packet denoting a meson bound state, such as $\vert B\,(\vec{p}, S_{B})\rangle$ at a specific momentum $\vec{p}$ and spin $S_{B}$, is formulated as follows~\cite{Nayak:2021djn,Nayak:2022gdo}
\begin{equation}
\big\vert B \left(\vec{p},S_{B}\right)\big\rangle = \hat{\Lambda}\left(\vec{p},S_{B}\right)\big\vert \left(\vec{p_b},\lambda_b\right);\left(\vec{p_d},\lambda_d\right)\big\rangle \,.
\label{eq:MesonWF}
\end{equation}
In the given expression, $\vert (\vec{p_b},\lambda_b);(\vec{p_d},\lambda_d)\rangle$ denotes the Fock space representation of the unbound quark and antiquark in a color-singlet configuration with their respective momentum and spin, and ${\hat \Lambda}(\vec {p},S_{B})$ represents an integral operator encapsulating the bound state characteristics of a meson. The same meson structure along with the corresponding momentum and spin configurations can be extended to related meson states such as $B_u$, $B_s$, $D$ and $D_s$.

Hadronic matrix elements encapsulate non-perturbative aspects of QCD and cannot be calculated directly from first principles. However, the matrix element describing a transition between hadrons with specific spin and parity, under the influence of a given operator can be expressed using a finite number of amplitudes corresponding to partial waves with well defined orbital angular momentum. Each amplitude takes the form of a tensor product involving external momenta, polarization vectors, and spinors, multiplied by an unknown hadronic quantity known as a form factor. These matrix elements can be written in various equivalent forms, each using a different linear combination of tensor structures, thereby establishing a basis for the form factors~\cite{Bernlochner:2021vlv}. For $B \to D\, (D^*)$ and $B_s \to D_s \,(D_s^*)$ SM transitions, the matrix elements are represented by two (four) independent form factors. For $0^- \to 0^-$ and $0^- \to 1^-$ type transitions, the vector and axial vector current matrix elements involving the form factors are defined as:
\begin{align}
    \langle{D\,(D_s)} |\bar c\,\gamma_{\mu}\,b|{B\,(B_s)}\rangle = (p+k)_{\mu} \, f_{+}(q^2) \,+\, q_{\mu}\,f_{-}(q^2),
    \label{eq:ff1}
\end{align}

\begin{align}
    \langle{D^*\,(D_s^*)} |\bar c\,\gamma_{\mu}\,b|{B\,(B_s)}\rangle = \frac{1}{M + m}\, \epsilon^{\sigma^+}\Big \{i\, \epsilon_{\mu\sigma\alpha\beta}\,(p + k)_{\alpha}\,q_{\beta}\,V(q^2)\Big\},
    \label{eq:ff2}
\end{align}

\begin{align}
    \label{eq:ff3}
    \langle{D^*\,(D_s^*)}| \bar c\,\gamma_{\mu}\,\gamma_{5}\,b|{B\,(B_s)}\rangle =& \frac{1}{M + m}\, \epsilon^{\sigma^+}\times\\\nonumber
    &\Big\{ g_{\mu \sigma}\, (p+k) \,q_{\mu}\, A_{0}(q^2)\, +\,\\\nonumber
    &(p+k)_{\mu}\, (p+k)_{\sigma}\,A_{+}(q^2)\,+\, \\\nonumber
    &q_{\mu}\,(p+k)_{\sigma}\,A_{-}(q^2)\Big\}.
\end{align}

Note that here $\langle{D\,(D_s)}| \bar c\,\gamma_{\mu}\gamma_{5}\,b|{B\,(B_s)}\rangle = 0$ because of angular momentum and parity conservation. The spectroscopy basis of form factors has been taken from the heavy quark symmetry (HQS)~\cite{Isgur:1991wq,Neubert:1993mb} considerations. 

The hadronic amplitude $\mathcal{H}_{\mu}$ in~\eqref{eq:TME} is derived by evaluating the overlap integral of the meson wavefunctions given in~\eqref{eq:MesonWF}. Using the wave packet description of the involved meson states, this leads to the standard expression for the $S$-matrix element, $S_{fi}$ describing the decay process,
 \begin{equation}
 	S_{fi}=(-{\cal M})\,\delta^{(4)}(p-k-k_{\tau}-k_{\nu_{\tau}})\,\frac{(\pi)}{\sqrt{E_{B}E_X}}\,\Pi_f .
 \end{equation} 
 \begin{widetext}
The hadronic amplitude ${\cal H}_\mu$ in the parent meson rest frame is obtained as:
\begin{equation}
	{\cal H}_\mu=\sqrt{\frac{ME_k}{N_{B}(0)N_X(\vec{k})}}\int\frac{d^3p_b}{\sqrt{E_{p_b}E_{k+p_b}}}\,{\cal G}_{B}(\vec{p_b},-\vec{p_d})\,{\cal G}_X(\vec{k}+\vec{p_b},-\vec{p_d})\,\langle S_X\vert J_\mu^h(0)\vert S_{B}\rangle. 
    \label{eq:HA}
\end{equation} 	  

Here, $E_{p_b}$ and $E_{k + p_b}$ denote the energies of the non-spectator quark in the parent and daughter mesons, respectively, while $\langle S_X \vert J_\mu^h(0) \vert S_{B} \rangle$ symbolically represents the spin matrix elements of the hadronic vector–axial vector current. For $0^-\to 0^-$ transitions, the axial-vector current yields no contribution. Consequently, the spin matrix elements arising from the non-zero components of the vector current take the form:
\begin{eqnarray}
	\langle S_X(\vec{k})\vert V_0\vert S_{B}(0)\rangle =\frac{(E_{p_b}+m_b)(E_{p_{c}}+m_{c})+|\vec{p_b}|^2}{\sqrt{(E_{p_b}+m_b)(E_{p_{c}}+m_{c})}}\,, 
	\langle S_X(\vec{k})\vert V_i\vert S_{B}(0)\rangle =\frac{(E_{p_b}+m_b)\,k_i}{\sqrt{(E_{p_b}+m_b)(E_{k+p_b}+m_{c})}}.
\end{eqnarray}	  
 Using the spin matrix elements derived above, the hadronic amplitudes in~\eqref{eq:HA} are equated to~\eqref{eq:ff1}, leading to the extraction of the form factors $f_+$ and $f_-$ for ($0^-\to 0^-$) transition in the form:

\begin{eqnarray}
 	  f_\pm (q^2)=\frac{1}{2M}\sqrt{\frac{ME_k}{N_{B}(0)N_X(\vec{k})}}\int d\vec{p_b}\,{\cal G}_{B}(\vec{p_b},-\vec{p_d})\,{\cal G}_X(\vec{k}+\vec{p_b},-\vec{p_d})\times \frac{(E_{p_b}+m_b)(E_{p_{c}}+m_{c})+|\vec{p_b}|^2\pm(E_{p_b}+m_b)(M\mp E_k)}{E_{p_b}E_{p_{c}}(E_{p_b}+m_b)(E_{p_{c}}+m_{c})}	\label{eq:fplusminus}
 	  \end{eqnarray}
      
 For $(0^-\to 1^-)$ transitions, the spin matrix elements associated with the vector and axial-vector currents are obtained independently in the following form:
\begin{eqnarray}
\label{eq:vectorFS2}
 \langle S_X(\vec{k},\hat{\epsilon^*})\vert V_0\vert S_{B}(0)\rangle =\,&&0 \,\,\,\,\text{and}\,\,\, \, \langle S_X(\vec{k},\hat{\epsilon^*})\vert V_i\vert S_{B}(0)\rangle=\frac{i(E_{p_b}+m_b)(\hat{\epsilon}^*\times \vec{k})_i}{\sqrt{(E_{p_b}+m_b)(E_{p_b+k}+m_{c})}},\\
 \langle S_X(\vec{k},\hat{\epsilon^*})\vert A_i\vert S_{B}(0)\rangle=&&\frac{(E_{p_b}+m_b)(E_{p_b+k}+m_{c})-\frac{|\vec{p_b}|^2}{3}}{\sqrt{(E_{p_b}+m_b)(E_{p_b+k}+m_{c})}}\,\,\,\,\text{and}\,\,\,\,
 \langle S_X(\vec{k},\hat{\epsilon^*})\vert A_0\vert S_{B}(0)\rangle=\frac{-(E_{p_b}+m_b)(\hat{\epsilon}^*. \vec{k})}{\sqrt{(E_{p_b}+m_b)(E_{p_b+k}+m_{c})}}.
 \label{eq:vectorFS}
 \end{eqnarray}
 	  
 Using the spin matrix elements~\eqref{eq:vectorFS2} -~\eqref{eq:vectorFS}, the hadronic amplitudes in~\eqref{eq:HA} are compared with the corresponding expressions~\eqref{eq:ff2} -~\eqref{eq:ff3}. This comparison yields the model expressions for the form factors $V(q^2),A_0(q^2),A_+(q^2)$ and $A_-(q^2)$ in the following form:
 \begin{eqnarray}
 V(q^2)=\frac{M+m}{2M}\sqrt{\frac{ME_k}{N_{B}(0)N_X(\vec{k})}}\int d\vec{p_b}\,{\cal G}_{B}(\vec{p_b},-\vec{p_d})\,{\cal G}_X(\vec{k}+\vec{p_b},-\vec{p_d}) \,\times \sqrt{\frac{(E_{p_b}+m_b)}{E_{p_b}E_{p_{c}}(E_{p_{c}}+m_{c})}},
 \label{eq:V0}
\end{eqnarray}

\begin{eqnarray}
 A_0(q^2)=\frac{1}{(M-m)}\sqrt{\frac{Mm}{N_{B}(0)N_X(\vec{k})}}\int d\vec{p_b}\,{\cal G}_{B}(\vec{p_b},-\vec{p_d})\,{\cal G}_X(\vec{k}+\vec{p_b},-\vec{p_d})\, \times \frac{(E_{p_b}+m_b)(E^0_{p_{c}}+m_{c})-\frac{|\vec{p_b}|^2}{3}}{\sqrt{E_{p_b}E_{p_{c}}(E_{p_b}+m_b)(E_{p_{c}}+m_{c})}},
 \label{eq:A0}
\end{eqnarray}
  	
\begin{center}
where, $E^0_{p_{c}}=\sqrt{|\vec{p}_{c}|^2+m^2_{c}}$
\end{center}

\begin{equation}
A_\pm(q^2)=\frac{-E_k(M+m)}{2M(M+2E_k)}\big[T\mp \frac{3(M\mp E_k)}{(E_k^2-m^2)}\big\{I-A_0(M-m)\big\}\big]\, ,T=J-(\frac{M-m}{E_k})A_0,
\label{eq:Aplusminus}
\end{equation}

\begin{eqnarray}
J=\sqrt{\frac{ME_k}{N_{B}(0)N_X(\vec{k})}}\int d\vec{p_b}\,{\cal G}_{B}(\vec{p_b},-\vec{p_d})\,{\cal G}_X(\vec{k}+\vec{p_b},-\vec{p_d})\, \times \sqrt{\frac{(E_{p_b}+m_b)}{E_{p_b}E_{p_{c}}(E_{p_{c}}+m_{c})}},
\end{eqnarray}
\begin{eqnarray}
I=\sqrt{\frac{ME_k}{N_{B}(0)N_X(\vec{k})}}\int d\vec{p_b}\,{\cal G}_{B}(\vec{p_b},-\vec{p_d})\,{\cal G}_X(\vec{k}+\vec{p_b},-\vec{p_d})\nonumber \, \times \biggl\{\frac{(E_{p_b}+m_b)(E^0_{p_{c}}+m_{c})-\frac{|\vec{p}_b|^2}{3}}{\sqrt{E_{p_b}E^0_{p_{c}}(E_{p_b}+m_b)(E^0_{p_{c}}+m_{c})}}\biggr\}.
\end{eqnarray}
\end{widetext}
With the relevant form factors thus obtained in terms of model quantities, the helicity amplitudes and hence the decay rates for $B \to D^{(*)}$ and $B_s \to D_s^{(*)}$ are evaluated in the following section and our predictions are listed in~\ref{sec:num_res}.

\subsection{Helicity Amplitudes}
\label{sub:helicity}
Using weak form factors from the covariant expansion of hadronic amplitudes, the angular decay distribution in  $q^2$ is determined. Here, $q = p - k = k_{\tau} + k_{\nu_{\tau}}$, and the decay distribution is calculated within the allowed kinematic range: $0 \leq q^2 \leq (M-m)^2$~\cite{Nayak:2021djn}
\begin{equation}
\frac{d\Gamma}{dq^2 d\cos\theta} = {\frac{{\cal G}_F}{(2\pi)^3}}|V_{cb}|^2\frac{(q^2-m_{\tau}^2)^2}{8M^2 q^2}|\vec{k}|{\cal L}^{\mu \sigma}{\cal H}_{\mu \sigma}\,.
\label{eq:10}
\end{equation}
In the above ${\cal L}^{\mu \sigma}$ and ${\cal H}_{\mu \sigma}$ represent the lepton and hadron correlation functions, respectively. $m_{\tau}$ denotes the mass of the $\tau$ lepton,  and $M$ represents the mass of the parent meson. Utilizing the completeness property, the lepton and hadron tensors in Eq.~\eqref{eq:10} are given as 
\begin{eqnarray}
{\cal L}^{\mu \sigma}{\cal H}_{\mu\sigma}              &=&{\cal L}_{\mu'\sigma'}g^{\mu'\mu}g^{\sigma'\sigma}{\cal H}_{\mu\sigma}\,,\nonumber\\
&=&{{\cal L}_{\mu'\sigma'}}{\epsilon^{\mu^{'}}}(m){\epsilon^{\mu^{\dagger}}}(m^{'}){g_{mm^{'}}}{\epsilon^{\sigma^{\dagger}}}(n){\epsilon^{\sigma^{'}}}(n^{'})g_{nn^{'}}{\cal H}_{\mu\sigma}\,,\nonumber\nonumber\\
&=&{L(m,n)}{g_{mm'}}{g_{nn'}}H({m'}{n'})\,.
\end{eqnarray}
For convenience, the lepton and hadron tensors are introduced in the space of helicity components
\begin{eqnarray}
L(m,n)&=&{\epsilon^{\mu}}(m){\epsilon^{\sigma ^\dagger}}(n){\cal L}_{\mu \sigma}\,,\nonumber\\
H(m,n)&=&{\epsilon^{\mu^\dagger}}(m){\epsilon^{\sigma}(n)}{\cal H}_{\mu \sigma}\,.
\end{eqnarray}
To facilitate the analysis of decay processes, physical observables are formulated in terms of helicity components, which offer a convenient and physically intuitive basis for describing angular distributions and polarization effects. As a result, helicity form factors are introduced by projecting the Lorentz-invariant form factors defined in Eqs.~\eqref{eq:ff1} -- \eqref{eq:ff3} onto specific helicity states of the final state particles. These form factors encapsulate the decay dynamics and are computed through overlap integrals of the initial and final meson wavefunctions as prescribed by the underlying model framework. The contraction of Lorentz indices as shown in Eq.~\eqref{eq:10} is performed with the helicity amplitudes following the formalism outlined in~\cite{Nayak:2021djn}, thereby connecting the covariant formulation with the helicity-based decomposition.

In the present analysis, we omit the azimuthal angular dependence $\chi$ associated with the lepton pair. Accordingly, the lepton tensor is integrated over the azimuthal angle, effectively removing any $\chi$ dependent contributions from the decay distribution. This simplification allows us to focus on the leading angular structure, yielding a differential decay distribution expressed solely in terms of the squared momentum transfer $q^2$ and the polar angle $\theta$ of the charged lepton. The resulting double differential distribution in the variables $(q^2, \cos\theta)$ is thus given by:

\begin{eqnarray}
\frac{d\Gamma}{d{q^2}\cos\theta}&=&\frac{3}{8}(1+\cos^2\theta)\frac{d\Gamma_U}{dq^2}+\frac{3}{4}\sin^2\theta \frac{d\Gamma_L}{dq^2}\nonumber\\
&\mp& \frac{3}{4} \cos\theta \frac{d\Gamma_P}{dq^2}+\frac{3}{4}\sin^2\theta\frac{d{\tilde{\Gamma}_U}}{dq^2}+\frac{3}{2}\cos^2\theta\frac{d{\tilde{\Gamma}_L}}{dq^2}\nonumber\\
&+&\frac{1}{2}\frac{d{\tilde{\Gamma}}_S}{dq^2}+3\cos\theta\frac{d{\tilde{\Gamma}}_{SL}}{dq^2}\,.
\label{eq:dcos}
\end{eqnarray} 
The upper and lower signs in~\eqref{eq:dcos} are related to the parity-violating term. Among the seven terms, four terms are recognized as tilde rates $\tilde{\Gamma}_i$, linked to the flip factor $m_{\tau}^2 / (2q^2)$, while the remaining terms are identified as $\Gamma_i$, which are independent of flip factor and they are related as:
\begin{equation}
\frac{d{\tilde{\Gamma}}_i}{dq^2}=\frac{m_{\tau}^2}{2q^2}\frac{d{\Gamma_i}}{dq^2}\,,
\end{equation}
where it is easy to notice that in the limit of vanishing lepton mass, the tilde rates do not contribute to the decay rate. While they can be disregarded for $e$ and $\mu$ modes, they are anticipated to have a significant impact on $\tau$-modes. Consequently, the tilde rates play a crucial role in assessing the lepton mass effects in the semileptonic decay modes. The differential partial helicity rates $d\Gamma_i / dq^2$ are defined by
\begin{equation}
\frac{d\Gamma_i}{dq^2}=\frac{{\cal G}_f^2}{(2\pi)^3}{\vert V_{cb}\vert}^2 \frac{({q^2}-{m_{\tau}^2})^2}{12{M^2}q^2}|\vec{k}|H_i\,.
\label{eq:dw}
\end{equation}
Here, $H_i$ ($i=U,L,P,S,SL$) denotes the conventional set of helicity structure functions, each defined as a specific linear combination of the helicity components of the hadronic tensor. These components are constructed from the bilinear products $H(m,n)=H_m H_n^\dagger$ where $H_m$ and $H_n$ correspond to the helicity amplitudes of the hadronic current.
\begin{eqnarray}
{H_U}&=&Re\left({H_+}{H_+^\dagger}\right)+Re\left({H_-}{H_-^\dagger}\right)\,,\nonumber\\
{H_L}&=&Re\left({H_0}{H_0^\dagger}\right)\,,\nonumber\\
{H_P}&=&Re\left({H_+}{H_+^\dagger}\right)-Re\left({H_-}{H_-^\dagger}\right)\,,\nonumber\\
{H_S}&=&3Re\left({H_t}{H_t^\dagger}\right)\,,\nonumber\\
{H_{SL}}&=&Re\left({H_t}{H_0^\dagger}\right)\,.
\end{eqnarray}
Here, $H_U$, $H_L$, $H_P$, $H_S$, and $H_{SL}$ represent unpolarized-transverse, longitudinal, parity-odd, scalar, and scalar-longitudinal interference functions, respectively.

We assume that the helicity amplitudes are real since the available $q^2$-range ($q^2\le(M-m)^2$) is below the physical threshold $q^2= (M+m)^2$. Therefore we neglect the angular terms multiplied by coefficients ${\rm Im}(H_iH_j^*), i\ne j^*$. Integrating over $\cos\theta$, one obtains the differential $q^2$ distribution. Finally integrating over $q^2$, one gets the total decay rate $\Gamma$ as the sum of the partial decay rates: $\Gamma_i$ ($i=U,L,P$) and $\tilde{\Gamma}_i$ ($i=U,L,S,SL$). 

Moving forward we explore a quantity of interest: the lepton polarization asymmetry for leptons in the final state. The polarization measurements of final state $\tau$ leptons have been proposed as a valuable tool for discerning BSM physics in various processes. Given the recent measurements of $P_{\tau}(D^*)$ and the fraction of $D^*$ longitudinal polarization $F_L^{D^*}$ by the Belle collaboration \cite{Belle:2016dyj, Belle:2019ewo} and LHCb~\cite{LHCb:2023ssl}, we extend our analysis and results in the subsequent section~\ref{sec:num_res} to include similar observables for $D$, $D^*$, $D_s$ and $D_s^*$ within the RIQM framework.
\begin{widetext}
\begin{figure}[hbt!]
\captionsetup{width=\textwidth}
\center
\begin{subfigure}{.5\textwidth}
  \center
\includegraphics[width=.8\linewidth]{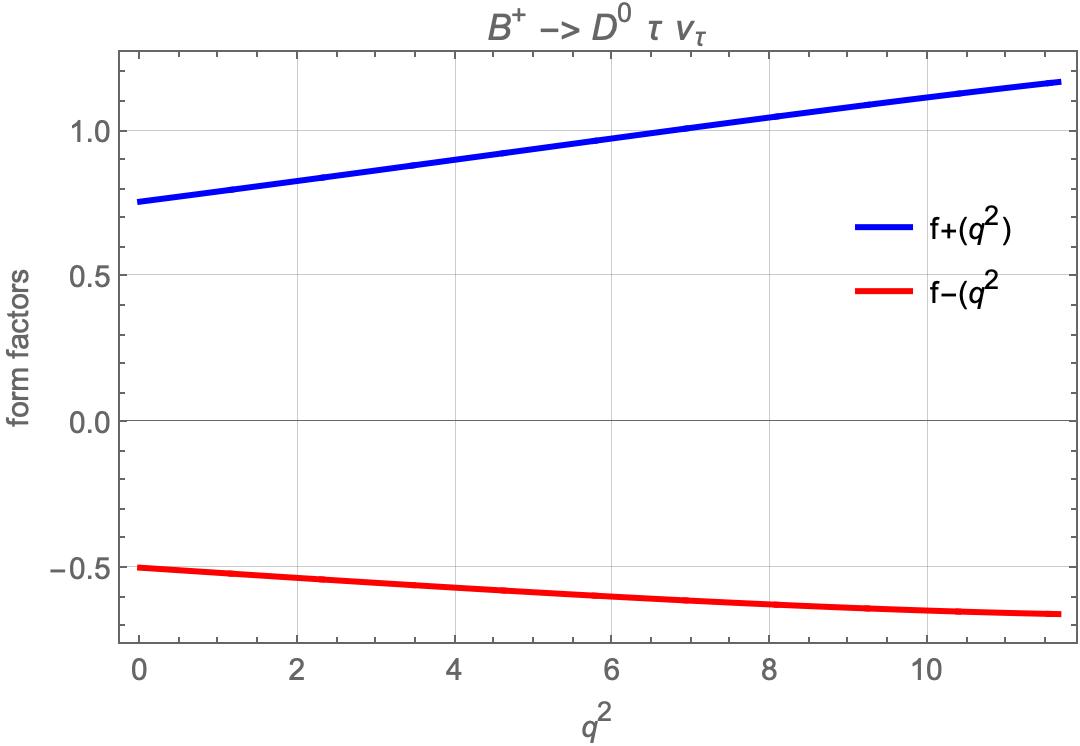}
\end{subfigure}%
\begin{subfigure}{.5\textwidth}
  \center
\includegraphics[width=.8\linewidth]{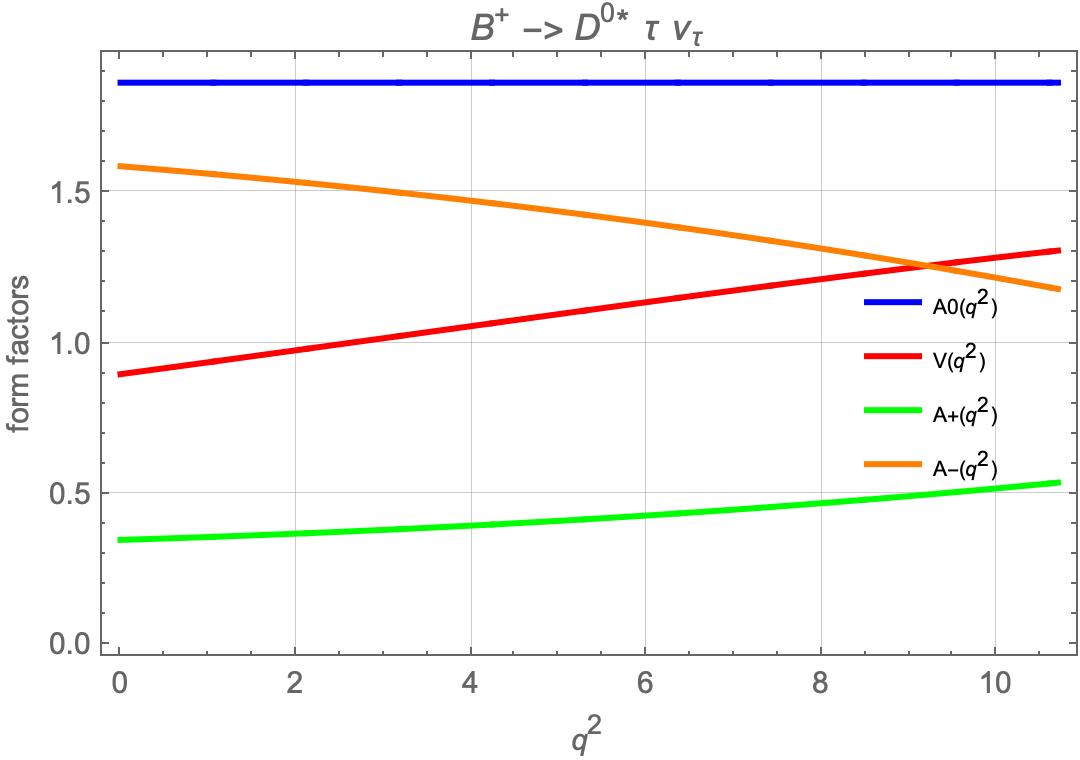}
\end{subfigure}
\begin{subfigure}{.5\textwidth}
  \center
\includegraphics[width=.8\linewidth]{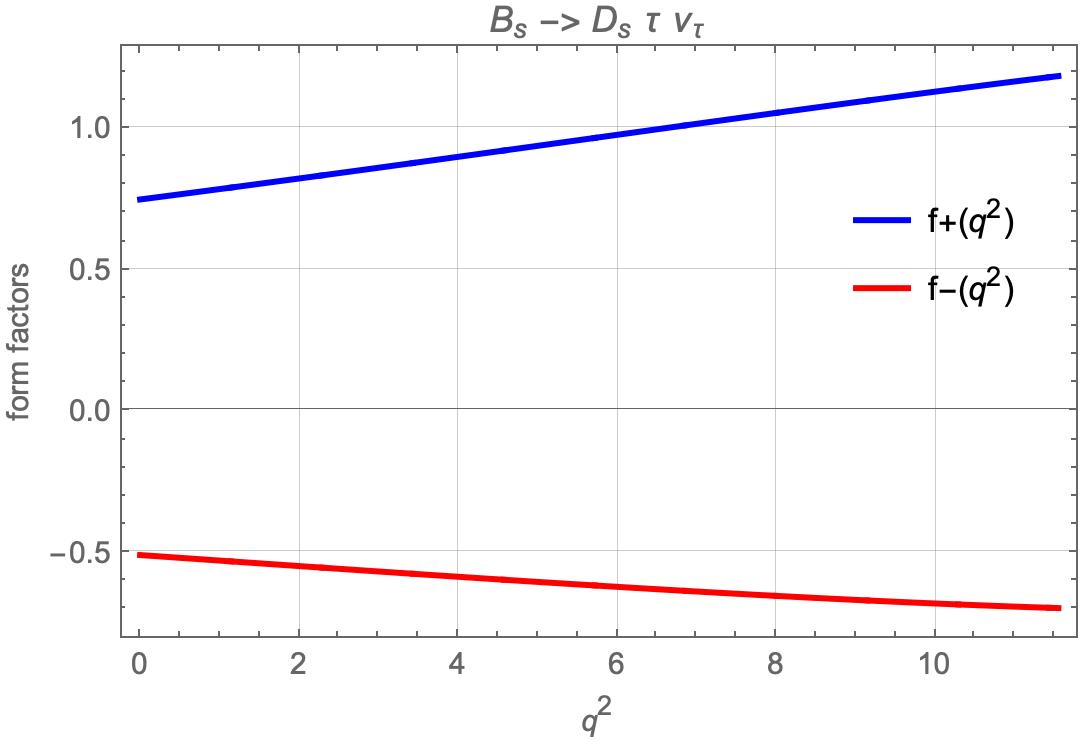}
\end{subfigure}%
\begin{subfigure}{.5\textwidth}
  \center
\includegraphics[width=.8\linewidth]{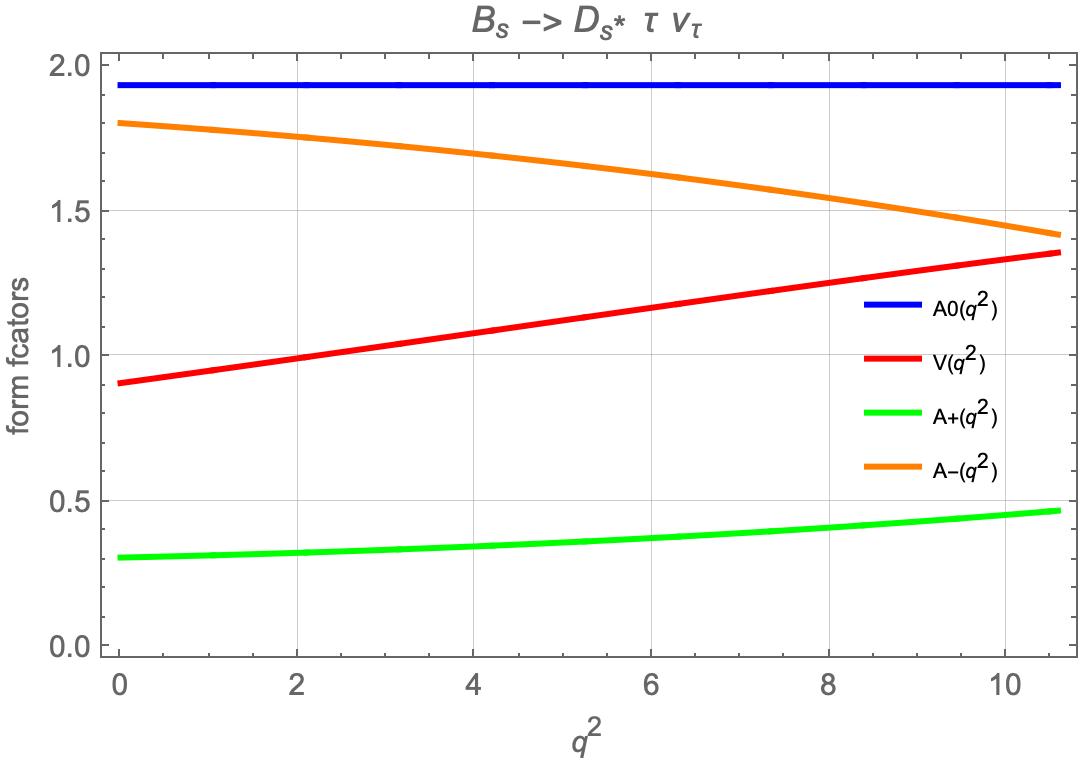}
\end{subfigure}
\caption{$q^2$ distribution spectra of the form factors for $B \to D(D^*) \tau \nu_{\tau}$ and $B_s \to D_s (D^*_s) \tau \nu_{\tau}$}
\label{fig:ffgraph}
\end{figure}
\end{widetext}

\section{Numerical results \& analysis}
\label{sec:num_res}
In this section, we present our numerical results and analysis of the exclusive semitauonic decays: $B \to D\,(D^{*})\, \tau \,\nu_{\tau}$ and $B_s \to D_s\,(D_s^{*}) \, \tau\, \nu_{\tau}$ in the RIQ model framework. We study the form factors, compute the branching fraction, polarization asymmetries and ratios of branching fraction in the muonic modes. To accurately describe these processes, we use different sets of model input parameters, for $B$ and $D$ mesons respectively, as outlined in~\eqref{Eq:IM}. The use of separate parameter sets reflects the variations in the binding energies of the constituent quark configurations and potential structures. 
\begin{eqnarray}
\text{B mesons} \, \to	(a, V_0)\,=\,&&(0.15\ {\rm GeV}^3,-0.5\ {\rm GeV})\,\nonumber\\
\text{D mesons} \, \to	(a, V_0)\,=\,&&(0.128\ {\rm GeV}^3,-0.5\ {\rm GeV})\,
    \label{Eq:IM}
\end{eqnarray}
We have taken the quark masses (in GeV) as,
\begin{eqnarray}
    m_{u/d}=0.305,\; m_s=0.510, \; m_c=1.500,\; m_b=4.700\,
\end{eqnarray}

which align closely with those in constituent quark model~\cite{Vijande:2004he,Capelo-Astudillo:2025fnp}. Using this set of constituent quark masses, and input parameters ($a, V_0$) we have reproduced the meson masses considered in this study, achieving close agreement of our predictions with the Particle Data Group (PDG)~\cite{ParticleDataGroup:2024cfk}. However, for computing the observables in the present study, we use the meson masses, lifetimes, and CKM matrix elements from the PDG~\cite{ParticleDataGroup:2024cfk} as they are precisely measured and will ensure a good comparison of our predictions with other theoretical approaches and experimental measurements. With the following input parameters, we proceed to first analyze the $q^2$ behaviour of the form factors in the accessible kinematic range in~\ref{sub:ff}. Then the predictions of relevant observables are presented in~\ref{sub:BFAO}.


%
\subsection{Weak decays form factors and \texorpdfstring{$q^2$} dependence}
\label{sub:ff}

The form factors are fundamental ingredients in describing the hadronic transition matrix elements involved in semileptonic decays. They encode the non-perturbative QCD dynamics of the parent and daughter mesons and are functions of the squared momentum transfer $q^2 = (p_B - p_D)^2$ which varies between $q^2_{min} = 0$ and $q^2_{max} = (m_B - m_D)^2$. In case of $0^- \to 0^-$ and $0^- \to 1^-$ transition,  the matrix elements of the vector and axial-vector currents can be decomposed into scalar, and vector form factors as outlined in~\eqref{eq:ff1}~-~\eqref{eq:ff3}. For $B \to D\,(D_s)$ the dominant contributions come from the form factor $f_+(q^2)$ and $f_{-}(q^2)$, which describe the vector current contributions, respectively. For $B \to D^*(D_s^{*})$, the decay involves more form factors: $V(q^2)$, $A_0(q^2)$, $A_+(q^2)$ and $A_-(q^2)$ corresponding to different components of the axial-vector and vector currents due to the spin-1 nature of the final state mesons. In our model framework the form factors are computed by evaluating the hadronic matrix elements using the constituent quark wavefunctions derived from the harmonic type confining potential. The model employs a mixed scalar-vector interaction, providing a Lorentz-invariant treatment of quark confinement effects. The meson bound-state wavefunctions, obtained as solutions of the Dirac equation with the specified confining potential, lead to the form factors through their overlap integrals, as presented in~\eqref{eq:fplusminus} and \eqref{eq:V0} - \eqref{eq:Aplusminus}. This covariant approach allows a smooth evaluation of the matrix elements across the full $q^2$ spectrum, which is especially important for exclusive transitions to study the recoil effects. The form factors at the maximum recoil point ($q^2 \to q^2_{min}\to0$) represent the scenario in which the final state meson has the largest possible three-momentum $|\vec{k}| = \frac{M^2 - m^2}{2M}$  in the rest frame of the decaying meson. In contrast, at the zero recoil point ($q^2 \to q^2_{max}$) the form factors reflect the overlap between the wave functions of the initial and final mesons. We provide our predicted values of the form factors in the limits $q^2 \to q^2_{min}$ and  $q^2 \to q^2_{max}$  as summarized in the Table~\ref{tab:FFP}. The form factors for both transitions $B \to D^{(*)}$ and $B_s \to D_s^{(*)}$ exhibit closely matching values throughout the entire kinematic range, as the effects of symmetry breaking are minimal — less than 10\%~\cite{Hu:2019bdf}.\\

Fig~\ref{fig:ffgraph} illustrates the calculated form factors as functions of $q^2$ for both $B \to D\, (D^*) \tau \nu_{\tau}$ and $B_s \to D_s \,(D_s^*) \tau \nu_{\tau}$ transitions. 
The form factor $f_+(q^2)$
(blue curve) exhibits a monotonically increasing behavior with respect to the momentum transfer $q^2$ reaching about $11$ GeV$^2$ across the kinematically allowed region. The form factor $f_{-}(q^2)$ (red curve) shows a negative slope, gradually decreasing with increasing $q^2$. This trend is consistent with expectations from HQET~\cite{Bernlochner:2017jka} and pole-dominance parametrization~\cite{Fan:2013kqa},where $f_+(q^2)$ generally dominates the decay amplitude, and contributes significantly in processes involving massive final state $\tau$ lepton where the phase space is reduced and shifted to large $q^2$ region if compared with electronic or muonic modes. A qualitative comparison with the approach used by CQM~\cite{Ivanov:2015tru,Ivanov:2016qtw,Ivanov:2017mrj} also confirms this picture, where the authors have computed the form factors from one-loop quark diagrams, ensuring full Lorentz covariance and confinement effects. A similar phenomenological behavior is observed in $B_s \to D_s$ case. The slightly larger values of $f_+(q^2)$ in this case, compared to the $B^+ \to D^0$ transition, reflect the differences in the internal quark dynamics and phase space of the $B_s \to D_s$ channel.

For the vector meson in the final states channels, we see $A_0(q^2)$ is constant, this is attributed to the fact that the form factor $A_0(q^2)$ is weakly dependent on $q^2$, almost independent in our model framework. This indicates that the contribution from the scalar part of the axial current (which contracts with the lepton current for longitudinal components) is relatively uniform. Physically, $A_0(q^2)$ contributes to the longitudinal polarization of the vector meson and can be less sensitive to recoil effects. Besides, $V(q^2)$ and $A_+(q^2)$ rises because the overlap of wavefunctions increases at large recoil. $A_-(q^2)$ often decreases due to suppression from kinematic factors or cancellations, controlling the longitudinal and transverse contributions, respectively. Overall the behaviour of our form factors are consistent with the CQM predictions~\cite{Ivanov:2016qtw}. The $q^2$ dependence spectra are also consistent with the general trends as predicted by (PQCD)~\cite{Hu:2019bdf} and (LQCD)~\cite{Harrison:2017fmw,Blossier:2021azl,Harrison:2021tol,Aoki:2023qpa} for $B \to D^*$  and $B_s \to D_s^{(*)}$ channel. The deviations fall within the theoretical uncertainties expected from model dependent treatments. 

If we compare our RIQ model parametrization with the results from BCL~\cite{Fan:2013qz,Boyd:1995sq} or z-expansion parametrization, the central values at low and moderate $q^2$  would show only minor deviations (within 5--10\%), but differences may grow near $q^2_{\max}$ due to the increasing sensitivity to the parametrization shape in that region. A key strength of our model lies in its capability to compute the decay form factors not just at a specific kinematic point, but across the entire $q^2$ range without relying on any supplementary assumptions or extrapolations. These form factors are input into the helicity amplitude framework to compute differential decay widths and angular observables. Their precise shapes directly influence predictions for polarization observables as detailed in the next section.

\begin{widetext}

\begin{table}[]
    \centering
    \begin{tabular}{|p{2cm}|p{2cm}|p{2cm}|p{2cm}|p{2cm}|}
    \hline
    \hline
    Form Factors  & $B \to D \tau \nu_{\tau}$ & $B \to D^* \tau \nu_{\tau}$ & $B_s \to D_s \tau \nu_{\tau}$ & $B_s \to D_s^* \tau \nu_{\tau}$\rule[-1ex]{0pt}{3ex} \\
    \hline
   
    \multirow{2}{4em}{$f_+(q^2)$}& $q^2_{min} = 0.87 $ & $-$ & $q^2_{min} = 0.74$  & $-$ \rule[-1ex]{0pt}{3ex} \\

    & $q^2_{max} = 1.16$ & & $q^2_{max} = 1.18$ & \rule[-1ex]{0pt}{3ex}\\
    \hline
    \multirow{2}{4em}{$f_-(q^2)$}
      & $q^2_{min} = -0.55$   & $-$ & $q^2_{min} = -0.51$   & $-$ \rule[-1ex]{0pt}{3ex} \\
      &$q^2_{max} = -0.65$ &&$q^2_{max} = -0.69$&\rule[-1ex]{0pt}{3ex}\\
    \hline
    \multirow{2}{4em}{$V(q^2)$}& & $q^2_{min} =0.89 $ &   & $q^2_{min} = 0.90$ \rule[-1ex]{0pt}{3ex}  \\

    &  & $q^2_{max} = 1.30 $&  & $q^2_{max} = 1.35$ \rule[-1ex]{0pt}{3ex}\\
    \hline
    \multirow{2}{4em}{$A_0(q^2)$}& & $q^2_{min} =1.86 $ &   & $q^2_{min} = 1.93$\rule[-1ex]{0pt}{3ex}  \\
    &  & $q^2_{max} = 1.86 $&  & $q^2_{max} = 1.93$ \rule[-1ex]{0pt}{3ex}\\
    \hline
        \multirow{2}{4em}{$A_+(q^2)$ }& & $q^2_{min} =0.34 $ &   & $q^2_{min} = 0.31$ \rule[-1ex]{0pt}{3ex} \\

    &  & $q^2_{max} = 0.53 $&  & $q^2_{max} = 0.46$ \rule[-1ex]{0pt}{3ex}\\
    \hline

      \multirow{2}{4em}{$A_-(q^2)$}& & $q^2_{min} =1.58 $ &   & $q^2_{min} = 1.80$ \rule[-1ex]{0pt}{3ex}  \\
 
    &  & $q^2_{max} = 1.17 $&  & $q^2_{max} = 1.42$ \rule[-1ex]{0pt}{3ex}\\
    
    \hline
    \hline
    \end{tabular}
    \caption{Results of form factors at $q^2 \to q^2_{min}$ and  $q^2 \to q^2_{max}$ predictions for $B$ and $B_s$ decays}
    \label{tab:FFP}
\end{table}
\end{widetext}
\begin{table*}[!htb]
\centering
\vspace{0.2cm}
\begin{tabular}{l|c| c c c| c c c c |c} \hline \hline
Channels & This work &  PQCD ~\cite{Hu:2019bdf} &  PQCD+Lattice~\cite{Hu:2019bdf} &LQCD~\cite{Harrison:2023dzh,MILC:2015uhg,McLean:2019qcx} & RQM~\cite{Faustov:2012mt}  & LCSR~\cite{Li:2009wq} & CQM~\cite{Ivanov:2015tru} &           HQET\cite{Fajfer:2012vx}  & PDG\cite{ParticleDataGroup:2024cfk} \rule[-2ex]{0pt}{5ex} \\ 
\hline
$B^+ \to D^0 \tau^+ \nu_\tau$ & $0.81_{-0.43}^{+0.44}$ & $0.86^{+0.34}_{-0.25}$ & $0.69^{+0.21}_{-0.17}$ & $0.65 \pm 0.04$  & $-$ & $-$ & $-$ & $ 0.66\pm 0.05 $ & $0.77\pm0.25$ \rule[-2ex]{0pt}{5ex}\\
\hline
$B^+ \to D^{0*} \tau^+ \nu_\tau $ & $1.96_{-0.07}^{+0.13}$ & $1.60^{+0.39}_{-0.37}$ & $1.34^{+0.26}_{-0.23}$ & $1.22 \pm 0.07$  & $-$ & $-$ & $-$ & $1.43\pm 0.05$ & $1.88\pm0.20$ \rule[-2ex]{0pt}{5ex}\\ 
\hline
$B^0 \to D^- \tau^+ \nu_\tau$ & $0.75_{-0.41}^{+0.40}$ & $0.82^{+0.33}_{-0.24}$ & $0.62^{+0.19}_{-0.14}$ & $-$  & $-$ & $-$ & 0.73 & $0.64\pm 0.05 $ & $0.99\pm0.21$ \rule[-2ex]{0pt}{5ex}\\
\hline
$B^0 \to D^{*-} \tau^+ \nu_\tau $ & $1.81_{-0.07}^{+0.12}$ &$1.53^{+0.37}_{-0.35}$ & $1.25^{+0.25}_{-0.21}$ & $-$ & $-$ & $-$ & 1.57 & $1.29\pm 0.06 $ & $1.45\pm0.10$ \rule[-2ex]{0pt}{5ex}\\
\hline
$B_s \to D_s \tau \nu_{\tau}$ & $0.76_{-0.49}^{+0.43}$ & $0.72^{+0.32}_{-0.23}$ & $0.63^{+0.17}_{-0.13}$ & 
$0.74 \pm 0.06$  & $0.62\pm0.05$
& $0.33^{+0.14}_{-0.11}$  & $-$ & $-$\rule[-2ex]{0pt}{5ex}  \\
\hline
$B_s \to D_s^* \tau \nu_{\tau}$ & $1.83_{-0.10}^{+0.11}$ & $1.45^{+0.46}_{-0.40}$ & $1.20^{+0.26}_{-0.23}$ & $1.25 \pm 0.05$ & $1.3\pm0.1$ & $-$ & $-$ & $-$ \rule[-2ex]{0pt}{5ex} \\
\hline\hline
\end{tabular}  
\caption{RIQM predictions for the branching fraction (in unit of $10^{-2}$ ) of the semitauonic decays}
\label{Tab:BR}
\end{table*}

\subsection{Branching Ratios and Angular Observables}
\label{sub:BFAO}

We integrate the differential decay rate~\eqref{eq:dw} to obtain the total branching fraction for all considered decay modes. Table \ref{Tab:BR} presents our predictions for the same (in units of $10^{-2}$) and compare them with results from PQCD~\cite{Hu:2019bdf}, LQCD~\cite{Harrison:2023dzh,MILC:2015uhg,McLean:2019qcx}, LCSR~\cite{Li:2009wq}, \textbf{CQM~\cite{Ivanov:2015tru}} and HQET~\cite{Fajfer:2012vx}. 
and other quark models~\cite{Faustov:2012mt} where available. Our uncertainties include $\pm 5\%$ variation due to changes in the model input parameters ($a$ and $V_0$), along with errors associated with the input parameters sourced from the PDG. We also compare our RIQ model results with the PDG values~\cite{ParticleDataGroup:2024cfk}, finding consistent agreement. Our values align well with other theoretical frameworks such as PQCD factorization formalism, and lie within experimental uncertainties. For the decays $B^+ \to D^0$ and $B_s \to D_s$ there are currently no direct LQCD predictions available for the absolute branching fractions. However, lattice collaborations have provided precise SM predictions for the ratios $\mathcal{R}_D$ and $\mathcal{R}_{D_s}$~\cite{MILC:2015uhg,McLean:2019qcx}. Using these theoretical predictions for $\mathcal{R}_D$ and $\mathcal{R}_{D_s}$ and the experimentally measured branching fractions for $B \to D\, (D_s) \,\ell\, \nu_{\ell}$ we estimated the branching fractions for LQCD approach in the tauonic modes to enable a qualitative comparison. Our results now show a consistency with the lattice predictions, agreeing well within the expected theoretical bounds in particular for $B_s \to D_s$ transition.


\begin{table*}[!htb]
\centering
\label{tab6}
\vspace{0.2cm}
\begin{tabular}{l| l| c c|cc  } \hline \hline
Observable      & Approach   &  $B^0\to D^- \tau^+\nu_\tau$ & $B^0_s\to D^-_s \tau^+ \nu_\tau $  & $B^0\to D^{*- } \tau^+ \nu_\tau $
& $B^0_s\to D_s^{*-} \tau^+ \nu_l $  \\ \hline
&This Work & $0.293$ & $0.291$ & $-0.53$ & $-0.53$ \\
&LQCD\cite{Harrison:2023dzh} & $-$ & $-$ & $-0.54(19)$ & $-0.53(91)$\\
&PQCD \cite{Hu:2019bdf}&  $0.32(1)$   &$0.31(1)$& $-0.54(1)$& $-0.54(1)$\\
$P_\tau(D_{(s)}^{(*)})$&PQCD+Lat \cite{Hu:2019bdf}.& $0.30(1)$& $0.30(1)$& $-0.53(1)$& $-0.53(1)$\\
&CQM\cite{Ivanov:2017mrj}&$0.33$ &$-$&$-0.50$ & $-$\\

&SM\cite{Bhattacharya:2018kig}&$0.335(4)$&$-$&$-0.491(25)$& $-$\\
&Belle\cite{Belle:2016dyj}&$-$&$-$&$-0.38\pm 0.51^{+0.21}_{-0.16}$&$-$\\   
\hline
&This Work& $-$ & $-$ & $0.48$ & $0.48$\\
&LQCD\cite{Harrison:2023dzh} &$-$ & $-$ & $0.39(24)$ & $0.42(12)$\\
&PQCD \cite{Hu:2019bdf}& $-$&$-$&$0.42(1)$& $0.42(1)$\\
&PQCD+Lat.\cite{Hu:2019bdf}& $-$&$-$&$0.43(1)$& $0.43(1)$\\
$F_{\rm L}(D_{(s)}^*)$&SM\cite{Bhattacharya:2018kig}& $-$&$-$&$0.457(10)$&$-$\\
&Belle\cite{Belle:2019ewo}&$-$&$-$&$0.60\pm 0.08\pm 0.04$&$-$\\
&LHCb\cite{LHCb:2023ssl}& $-$ & $-$ & $0.41 \pm 0.06 \pm 0.03$  & $-$\\
\hline\hline
\end{tabular}
\caption{Comparison of RIQM predictions for   $P_\tau(D_{(s)}^{(*)})$,   $F_L(D_{(s)}^*)$
with other SM approaches \& experimental measurements}
\label{Tab:PA}
\end{table*}

One key observation is that for channels involving $J^P = 1^-$ mesons in the final states, our model tends to predict slightly higher branching fractions than those involving $J^P = 0^+$ in the final states. This enhancement arises naturally within our model due to the treatment of relativistic effects and spin structure in the wavefunctions, which enhance the form factors in the vector channel more prominently. Despite this mild enhancement, the predictions remain well within the expected range and align closely with the PDG values and lattice-based results where available especially for $B_s \to D_s^* \tau \nu_{\tau}$ mode. This close correspondence between $B$ and $B_s$ decays indicates that the underlying SU(3) flavor symmetry is only mildly broken in these processes, with deviations typically remaining below the 10\% level~\cite{Bordone:2019guc}. Such small symmetry-breaking effects suggest that the hadronic dynamics governing these decays are largely universal thereby validating the use of symmetry based approximations in theoretical predictions. Thus the overall agreement of RIQ model predictions reinforces the validity of our approach in studying both pseudoscalar and vector transitions involving the $B$ and $B_s$ systems and can be extended to other systems in cases where lattice data and alternative theoretical methods are unavailable.

We further utilize our form factors to calculate two key angular asymmetries of the decays: the longitudinal $\tau$ polarization $P_{\tau}\,$ of the final states, and the longitudinal polarization fraction of the vector meson $F_L$. The integrated observables related to these quantities
are defined in~\ref{sec:physical_obs} with the numerators and denominators integrated over the whole $q^2$ independently.  In Table~\ref{Tab:PA} we report our results for these observables which show good agreement with the approaches, including PQCD~\cite{Hu:2019bdf} and the CQM~\cite{Ivanov:2017mrj} and other SM based approaches. Our prediction for $P_{\tau} (D^*) = -0.53$ matches well with Belle experiment~\cite{Belle:2016dyj}. Although there is a large statistical uncertainty on the experimental measurement. Similarly we obtain $F_L (D^*) = 0.48$ which lies within 
$2\sigma$ from the LHCb and Belle average~\cite{Belle:2019ewo,LHCb:2023ssl}, indicating statistical consistency with the experimental results. These results strengthen the credibility of our model’s predictions in probing the polarization observables as well.   



The ratios of the branching fractions of the exclusive $B \to D_{s}^{(*)-} \mu^+ \nu_{\mu}$ relative to those of exclusive $B^0 \to D^{(*)-} \mu^+ \nu_{\mu}$ decays are recently measured by LHCb \cite{LHCb:2020cyw}:
    
\begin{align}
    \mathcal{R} = \frac{\mathcal{B}(B_s \to D_{s}^{-} \mu^+ \nu_{\mu})}{\mathcal{B}(B^0 \to D^{-} \mu^+ \nu_{\mu})} = 1.09 \pm 0.05 \, (stat)&\pm 0.06\, (syst)\nonumber\\
    &\pm 0.05 \, (ext)\nonumber
\end{align}

\begin{align}
    \mathcal{R^*} = \frac{\mathcal{B}(B_s \to D_{s}^{(*)-} \mu^+ \nu_{\mu})}{\mathcal{B}(B^0 \to D^{(*)-} \mu^+ \nu_{\mu})} = 1.06 \pm 0.05 \, (stat) &\pm 0.07\, (syst)\nonumber\\ &\pm 0.05 \, (ext)\nonumber
\end{align}

These ratios provide a clean test of QCD and SU(3) symmetry and their proximity to $\sim 1$ suggests that semileptonic decays of $B$ and $B_s$ are well-understood. No anomalies or tensions are seen — consistent with the SM. Accordingly, we have computed these observables and obtained our values: $\mathcal{R} \sim 1.02$ and $\mathcal{R^*} \sim 1.03$ for muonic modes. Our results are compatible with the LHCb results at $0.1\sigma$. Similarly we constructed the corresponding ratios for the $\tau$ modes as well for the sake of completeness, obtaining:  $\mathcal{R} \sim 1.02$ and  $\mathcal{R^*} \sim 1.01$, which are again consistent with the above SM principles, reinforcing the robustness of the model framework in describing the decay dynamics in muonic and tauonic modes within the concepts of SM.

\section{Summary and outlook}
\label{sec:conclusions}
Motivated by the progressive measurements in refining the tensions in lepton flavor universality and recent experimental anomalies in $B$ decays, in the present study we have investigated the semitauonic decay modes: $B \to D \,(D^*)\, \tau\, \nu_{\tau}$ and $B_s \to D_s\, (D_s^*)\, \tau \,\nu_{\tau}$ using the \emph{Relativistic Independent Quark Model}. We computed the relevant form factors using a covariant potential model with harmonic confinement and evaluated a wide range of observables, including branching fractions, polarization asymmetries and clean test of QCD ratios. Our predictions for the branching fraction show good agreement with SM expectations and PDG values and lie within the current experimental bounds corroborating SU(3) flavor symmetry limit in $B$ and $B_s$ decays. In particular, the predicted values of $P_{\tau}$ and $F_L$ are consistent with the Belle and LHCb measurements, underscoring the reliability of the RIQM framework in capturing key features of $B$ semileptonic decays to charm states in tauonic modes. The calculated $q^2$ dependence form factors exhibit physically consistent behavior and adhere to theoretical constraints such as HQET normalization and endpoint scaling laws. We have also validated our predictions through comparisons with results from PQCD, LQCD, and other model approaches. 

Looking forward, the increasing precision at LHCb and Belle II along with the forthcoming Run3 data and possible future super-$B$ factories, will provide stringent tests for these observables. Our model-driven predictions provide an invaluable supplementary theoretical contribution to strengthen the understanding of SM dynamics in the realm of semileptonic $B$ decays. Even if NP exists beyond the reach of direct detection, the $B$ mesons decays will always remain a crucial tool for deciphering the SM physics. Continued synergy between phenomenological modeling, lattice simulations, and experimental measurements, will be critical in resolving the anomalies and advancing the potential discovery of $B$, $B_s$ and $B_c$ channels in the near future.
\begin{acknowledgments}
S.P. and L.N. are grateful for substantial contribution from Profs. N. Barik, P. C. Dash, and S. Kar in the development of the RIQ model framework. Authors also acknowledge NISER, Department of Atomic Energy, India, for financial support.
\end{acknowledgments}
\bibliography{bibliography}{}
\bibliographystyle{utphys}
\end{document}